\documentclass[letterpaper, 10 pt, conference]{ieeeconf}  

\IEEEoverridecommandlockouts                              
\usepackage[utf8]{inputenc}
\usepackage[T1]{fontenc}
\usepackage{multirow}
\usepackage[table,xcdraw]{xcolor}
\usepackage{colortbl}
\usepackage{graphicx}
\usepackage[normalem]{ulem}
\useunder{\uline}{\ul}{}
\usepackage{longtable}
\usepackage{cite}
\usepackage{booktabs}
\usepackage{threeparttable}

\title{\LARGE \bf
AI Thrust: Ranking Emerging Powers for Tech Startup Investment in Latin America
}

\author{ \parbox{3 in}{\centering Abraham Ramos Torres*
         \thanks{*This work was supported by the Accel AI Institute and produced on behalf of the LatinX in AI (LXAI) Organization.}\\
         Central University of Venezuela\\
         Caracas, Venezuela\\
         {\tt\small abraham@accel.ai}}
         \hspace*{ 0.5 in}
         \parbox{3 in}{ \centering Laura N Montoya* \\
         Accel AI Institute\\
         San Francisco, CA, USA\\
         {\tt\small laura@accel.ai}}
}

\usepackage[utf8]{inputenc}

\begin{document}

\maketitle

\begin{abstract}

Artificial intelligence (AI) is rapidly transforming the global economy, and Latin America is no exception. In recent years, there has been a growing interest in AI development and implementation in the region. This paper presents a ranking of Latin American (LATAM) countries based on their potential to become emerging powers in AI. The ranking is based on three pillars: infrastructure, education, and finance. Infrastructure is measured by the availability of electricity, high-speed internet, the quality of telecommunications networks, and the availability of supercomputers. Education is measured by the quality of education and the research status. Finance is measured by the cost of investments, history of investments, economic metrics, and current implementation of AI.

While Brazil, Chile, and Mexico have established themselves as major players in the AI industry in Latin America, our ranking demonstrates the new emerging powers in the region. According to the results, Argentina, Colombia, Uruguay, Costa Rica, and Ecuador are leading as new emerging powers in AI in Latin America. These countries have strong education systems, well-developed infrastructure, and growing financial resources. The ranking provides a useful tool for policymakers, investors, and businesses interested in AI development in Latin America. It can help to identify emerging LATAM countries with the greatest potential for AI growth and success.


\end{abstract}

\section{INTRODUCTION}

Artificial intelligence (AI) is a tool that can benefit most economic sectors including healthcare, education, environmental sustainability, and quality of life. In Latin America, a region, whose factors present various difficulties that hinder the advancement of its citizens, investment in drivers of this technology as a tool to further economic growth is crucial. 

The economic powers of the world have high investment, implementation, and strategies for artificial intelligence. The United States of America and China, are countries that support research into many uses of AI and resort to investments aimed at improving the infrastructure for its application and development\cite{Cesareo2023TheIndex}. AI development in these countries does not come exclusively from their governments, but also from private investment. The United States has some of the largest technology companies in the world, which count on the training of new talent, the production of new jobs, and technological development. In 2023 the size of the labor market in the United States for positions in the IT field neared 4.9 million employees\cite{JancoAssociates2023ItData}. On the other hand, AI could generate more than 20 new job roles in companies, such as AI Trainer Operator, AI Integration Specialist, AI Ethicist, and many others\cite{ExpertPanel202320Create}. 

The use of AI in the present, near, and long-term future was detailed by the economic sector in a recent publication by PWC\cite{2017SizingCapitalise}. While the use of AI in healthcare is presently focused primarily on health insurance operations, soon it will assist with the creation of virtual medicines and data-driven diagnoses, eventually yielding robotic doctors that manage diagnoses and treatment. In the automotive industry, AI is used for automated drive assistance. Soon, it will manufacture and maintain automated parts on demand, monitor predictions, and keep vehicles maintained. In financial services, AI is used for robotic advice, insurance underwriting, and process automation. Shortly, it will be used to design products based on customer preferences. In transportation and logistics, AI is used for automated picking in warehouses and will soon be used for traffic control and eventually, autonomous transportation and delivery. In technology, communication, and entertainment, AI can automate recommendations based on customer preferences. In the future, it will be able to automate telemarketing and even create content. In retail, AI is used for smart energy inventory management and will soon be able to produce goods and services based on automated market analysis and signals. In energy, AI is used for smart energy metering and will soon be used for optimizing energy management and creating new renewable technologies. Finally, in manufacturing, AI is used to automate production processes and will soon be used for predictive scheduling and automation in the supply chain and prospective analytics in product design.

Traditional economic sectors, such as manufacturing real GDP can benefit from the implementation of artificial intelligence. Robotic process automation (RPA), in conjunction with AI, can lead to an improvement in the development of goods and services, making them less expensive and more productive and efficient\cite{Dilmegani2024Top2024}. AI is being used by banks for customer service, fraud detection, financial advice, and banking processes. It can analyze data in real-time to provide personalized consultancies and prevent computer attacks\cite{Gupta2023ArtificialAi}. For infrastructure, the investment and implementation of AI both in administrative areas and practical tasks can increase the efficiency of results by reducing costs and increasing worker safety and effectiveness. Artificial intelligence can provide safer work when implemented in different ways, such as 3D printing, data analysis to predict potential accidents, and project monitoring and analysis. Overall, AI will be able to replace the 4 D’s of work, dull, dirty, dangerous, and difficult.

In Latin America, the investment potential is significant, even in areas where economic conditions are not favorable at first glance. When we investigate aspects needed for technological advancement and scalable growth in each of the countries in the region, we can observe that startups have huge economic potential, and even more so when they apply artificial intelligence in their operations or products. The following are some of the most important reasons why investment in artificial intelligence in Latin American countries can be favorable for the country and its citizens.

AI can have a significant impact on the economy of Latin American countries, it is expected that by 2030, AI will increase the region’s Real GDP by 5.4\% (0.5 trillion dollars)\cite{Dr.AnandS.RaoSizingCapitalise}. When investing in emerging markets such as Latin America, it is crucial to consider various economic factors such as the status of infrastructure, economy, and academic development. These factors can help determine the potential for profitable investment opportunities, especially in the field of AI. As Latin America's AI potential continues to grow, it is important to identify which countries offer the most promising prospects for investors. This research builds on prior findings on behalf of the Accel AI Institute and the LatinX in AI (LXAI) Organization, titled 'Government AI Readiness Meta-Analysis for Latin America And The Caribbean'\cite{Montoya2019GovernmentCaribbean}, strengthening insights for economic growth and external investment opportunities. Ultimately, we developed a ranking to help answer the question of which countries in Latin America would investing in AI, yield the highest profitability with the quickest return.

\section{Methodology}

Our methodology for assessing the potential for AI Investment in each country included the development of a ranking that accounts for the primary indicators under the infrastructure, education, and finance elements required to successfully launch and scale AI companies in each country.  

\subsection{Elements and Indicators}

Tables I, II, and III summarize the elements and indicators used for the ranking.

\begin{table}[ht]
\centering
\caption{Infrastructure Elements for AI Investment Ranking}
\label{tab:Metholodology-Infra}
\resizebox{\columnwidth}{!}{%
\begin{tabular}{p{0.17\linewidth} p{0.37\linewidth} p{0.30\linewidth} p{0.16\linewidth}}
\toprule
\multicolumn{4}{c}{Infrastructure Pillar} \\ \midrule
Element & Description & Indicator & Source \\ \midrule
Electricity & Does the country have access to electricity? & Access to Electricity (\%) & The World Bank\cite{WorldBankOpenData2023AccessPopulation} \\ \midrule
\multirow{5}{*}{Internet} & How is the internet distributed? & Access to Internet (\%) & World Bank Group\cite{WorldBankGroup2022Internet2021} \\ \cmidrule(l){3-4} 
 &  & Urban Internet Penetration & World Bank Group\cite{WorldBankGroup2022Internet2021} \\ \cmidrule(l){3-4} 
 &  & Rural Internet Penetration & World Bank Group\cite{WorldBankGroup2022Internet2021} \\ \cmidrule(l){3-4} 
 &  & Internet Speed (Fixed Broadband) & Speedtest Global Index\cite{Speedtest2023InternetWorld} \\ \cmidrule(l){3-4} 
 &  & Internet Speed (Mobile) & Speedtest Global Index \\ \midrule
Supercomputers & What is the existing supercomputer capacity within the country? & HPC Systems Availability & RISC2\cite{Hafner2021RISC2LATAM} \\ \bottomrule
\end{tabular}%
}
\end{table}

\begin{table}[ht]
\centering
\caption{Educational Elements for AI Investment Ranking}
\label{tab:Methodology-Education}
\resizebox{\columnwidth}{!}{%
\begin{tabular}{p{0.17\linewidth} p{0.37\linewidth} p{0.30\linewidth} p{0.16\linewidth}}
\hline
\multicolumn{4}{c}{Educational Pillar} \\ \hline
Element & Description & Indicator & Source \\ \hline
\multirow{3}{*}{Education} & \multirow{3}{*}{Tertiary Education Penetration?} & 25-34 years old (\%) Population with tertiary education & OECD\cite{OECDPopulationEducation} \\ \cline{3-4} 
 &  & 55-64 years old (\%) Population with tertiary education & OECD\cite{OECDPopulationEducation} \\ \cline{3-4} 
 &  & Number of Universities in Top 100 of Latin America & Quacquarelli Symonds Limited\cite{QSQuacquarelliSymonds2023QS2023} \\ \hline
\multirow{2}{*}{Research} & \multirow{2}{*}{Research Articles and Patents?} & Articles Published in the AI Field & ETO @ CSET\cite{CenterofSecurityandEmergingTechnology2023CountryIntelligence} \\ \cline{3-4} 
 &  & Patents Granted in the AI Field & ETO @ CSET\cite{CenterofSecurityandEmergingTechnology2023CountryIntelligence} \\ \hline
\end{tabular}%
}
\end{table}

\begin{table}[h]
\centering
\caption{FINANCIAL ELEMENTS FOR AI INVESTMENT RANKING}
\label{tab:method-finance}
\resizebox{\columnwidth}{!}{%
\begin{tabular}{p{0.17\linewidth} p{0.37\linewidth} p{0.30\linewidth} p{0.16\linewidth}}
\hline
\multicolumn{4}{c}{Finance Pillar} \\ \hline
Element & Description & Indicator & Source \\ \hline
\multirow{5}{*}{Startups} & \multirow{5}{*}{Startup Penetration?} & Startups and other Privately-held AI companies headquartered in the country? & ETO @ CSET\cite{CenterofSecurityandEmergingTechnology2023CountryIntelligence} \\ \cline{3-4} 
 &  & Startups by HQ (\%) & SFJD\cite{Campos2021LatinStartups} \\ \cline{3-4} 
 &  & Startups By Sector (\%) & LAVCA\cite{LAVCA2019LatinDIRECTORY} \\ \cline{3-4} 
 &  & Startups with Female Executives (\%) & LAVCA\cite{LAVCA2019GenderTech/Startups} \\ \cline{3-4} 
 &  & Startups with Female Investors (\%) & LAVCA\cite{LAVCA2019GenderTech/Startups} \\ \hline
\multirow{2}{*}{CS Salaries} & \multirow{2}{*}{What is the cost of tech labor?} & \multirow{2}{*}{Median CS Salary} & Glassdoor\cite{Glassdoor2023Salary:2023} \\ \cline{4-4} 
 &  &  & Stack Overflow\cite{StackOverflow20222022Survey} \\ \hline
AI Strategies & Are there AI strategies? & AI Government Strategy & OECD\cite{OCEDiLibraryLACStrategies} \\ \hline
\multirow{8}{*}{Investments} & \multirow{8}{*}{How much AI investment?} & Total incoming investments (millions USD) & ETO @ CSET\cite{CenterofSecurityandEmergingTechnology2023CountryIntelligence} \\ \cline{3-4} 
 &  & Incoming Investments & ETO @ CSET\cite{CenterofSecurityandEmergingTechnology2023CountryIntelligence} \\ \cline{3-4} 
 &  & Global Corporate Investment in AI (Billions USD) & Stanford University\cite{Human-CenteredArtificialIntelligence2023Artificial2023} \\ \cline{3-4} 
 &  & 2019-2020 Billions USD Invested Tech & LAVCA\cite{LAVCA2021LAVCAsAmerica}\cite{LAVCA2020LAVCAsAmerica} \\
 &  &  &  \\ \cline{3-4} 
 &  & 2019-2020 Number of Tech Investment Deals & LAVCA\cite{LAVCA2021LAVCAsAmerica}\cite{LAVCA2020LAVCAsAmerica} \\
 &  &  &  \\ \cline{3-4} 
 &  & Total Capital Invested By Top Sectors 2020 & LAVCA\cite{LAVCA2021LAVCAsAmerica} \\ \hline
\multirow{6}{*}{Economics} & \multirow{6}{*}{Country Economic Status?} & 2020 GDP (Millions) & The World Bank\cite{WorldBankGDPCaribbean} \\ \cline{3-4} 
 &  & 2021 GDP (Millions) & The World Bank\cite{WorldBankGDPCaribbean} \\ \cline{3-4} 
 &  & 2022 GDP per capita, PPP & The Global Economy\cite{TheGlobalEconomyGDPRankings} \\ \cline{3-4} 
 &  & Real GDP 2023 Projection (Annual \% Change) & IMF\cite{InternationalMonetaryFund2023WorldRecovery} \\ \cline{3-4} 
 &  & Real GDP 2024 Projection (Annual \% Change) & IMF\cite{InternationalMonetaryFund2023WorldRecovery} \\ \cline{3-4} 
 &  & Real GDP 2028 Projection (Annual \% Change) & IMF\cite{InternationalMonetaryFund2023WorldRecovery} \\ \hline
\multirow{3}{*}{AI Adoption} & AI Implementation across organizations? & Organizations implementing AI & IDC\cite{IDC2022AnaliticaDatos} \\ \cline{3-4} 
 &  & Survey of Organizations Who have Adopted AI in at Least One Function, 2017–2022 & McKinsey\cite{Chui2022TheReview} \\ \cline{3-4} 
 &  & AI Adoption by Organizations in LATAM between 2021 and 2022 & Stanford University\cite{Human-CenteredArtificialIntelligence2023Artificial2023} \\ \hline
\end{tabular}%
}
\end{table}

\subsection{Missing Values}

For countries with missing data, the ‘NA’ value was used to calculate the final scores (explained in the Calculating Scores section).

\subsection{Calculating Scores}

\subsubsection{Normalization}

All scores were Z-score normalized using the following formula:
\begin{equation}
    z = (x -  \mu ) / ( \sigma )
\end{equation}

Where: 

z is the normalized value.

x is the original value.

\( \mu \) is the media of the distribution.

\( \sigma \) is the standard deviation of the distribution.

All values used for normalization were original. Scores were calculated using the normalized formula based on media of Latin American country data.

Z-score normalization is a method of standardization that transforms the values of a feature by subtracting the mean (\( \mu \)) and dividing by the standard deviation (\( \sigma \)). This results in a dataset with a mean of zero and a standard deviation of one. 

This method was selected because it is less affected by outliers, which can skew the range and scaling of values in other methods. This method preserves the original distribution shape of values and allows for easy comparison regarding the number of standard deviations from the mean.

The value NA is commonly used to represent missing data. It acts as a placeholder for missing data and helps the formula recognize and handle the missing values correctly. NA stands for "not available" and can be used for both character and numeric data. Using NA allows the formula to ignore missing values during calculations or replace them with other methods.

\subsubsection{Limitations}

It is important to note that the data we included in our research is from 2019 to 2023, and the data from 2019 to 2020 may not accurately reflect the current situation in each country. Therefore, we urge caution when interpreting the results and when comparing them with more recent data from 2021 to 2023. 

Additionally, we encountered challenges in obtaining data for some countries or regions, especially those with low internet penetration or political instability. As a result, our data may not be representative of the entire region, and we may have missed relevant aspects or trends related to our research question. To mitigate this issue, we used various data sources, listed in Table I, and cross-checked them for consistency and validity.

\subsubsection{Ranking}

The ranking was developed using indicators that allowed for the calculation of individual scores for each country—indicators that provided general information for all Latin American countries were not considered.

Implementing artificial intelligence correctly requires following a series of procedures while taking into account various factors such as infrastructure, education level and publications, and finances of the country or territory where it will be implemented. In Latin America, these factors can vary significantly in the region's countries, making Latin America attractive to investors by having many alternatives where AI can be implemented.

\subsection{Infrastructure} 

Successful implementation of artificial intelligence in Latin America requires infrastructure access including electricity, internet availability, and distribution, fast internet speed, and the availability of supercomputers in both urban and rural areas\cite{Tortoise2021TheMethodology}. These factors are the most important for developing artificial intelligence (AI) solutions since the significant amount of power to operate, the essence of the data for creating unbiased models, the importance of speed when transferring large data sets, and collaboration between AI researchers and developers, and the role of the high computational power needed to implement the AI.

\subsection{Education and Research}

To achieve the best possible implementation of artificial intelligence, it is crucial to approach, develop, and plan it with the help of experts in the field of computer science who can carry out large-scale projects. Therefore, education and research on artificial intelligence are considered essential factors for companies, educational institutions, and government institutions.

\subsection{Finance}

Companies and countries must consider financial and economic factors when making decisions about investing in artificial intelligence. it's essential to ensure that the investment in AI is financially feasible and that the returns on investment are significant. Moreover, companies and countries must seek the potential to improve productivity and profitability through applying strategies based on the current economic status of the country. Risk management must be taken into account while investing in AI, and companies and countries must be aware of the potential risks involved. The possibility of innovation is another critical factor that companies and countries must consider. Investing in AI can lead to new opportunities for innovation, which can boost productivity and profitability. Understanding the competition in a specific location is essential for successful AI implementation. Companies and countries must consider the AI initiatives of their competitors and the potential impact on their business operations\cite{Szczepanski2019EconomicAI}. 

\section{RESULTS}

\begin{table}[ht]
\centering
\caption{LATIN AMERICAN AI INVESTMENT POTENTIAL RANKING}
\label{tab:ranking}
\resizebox{\columnwidth}{!}{%
\begin{tabular}{lrrrr}
\hline
 & \multicolumn{3}{c}{\textbf{ AI Investment Factors}} & \multicolumn{1}{l}{} \\ \hline
\textbf{Country} & \multicolumn{1}{l}{\textbf{Infrastructure}} & \multicolumn{1}{l}{\textbf{Educational}} & \multicolumn{1}{l}{\textbf{Financial}} & \multicolumn{1}{l}{\textbf{Final Ranking}} \\ \hline
Brazil & 1 & 1 & 1 & 1 \\ \hline
Mexico & 6 & 2 & 2 & 2 \\ \hline
Chile & 2 & 3 & 3 & 3 \\ \hline
Argentina & 3 & 4 & 5 & 4 \\ \hline
Colombia & 8 & 5 & 4 & 5 \\ \hline
Uruguay & 5 & 10 & 9 & 6 \\ \hline
Dominican Republic & 10 & 19 & 6 & 7 \\ \hline
Costa Rica & 7 & 6 & 8 & 8 \\ \hline
Ecuador & 4 & 8 & 11 & 9 \\ \hline
Panama & 12 & 12 & 7 & 10 \\ \hline
Peru & 15 & 7 & 10 & 11 \\ \hline
Venezuela & 13 & 9 & 13 & 12 \\ \hline
Paraguay & 9 & 13 & 15 & 13 \\ \hline
El Salvador & 11 & 17 & 18 & 14 \\ \hline
Cuba & 14 & 11 & 17 & 15 \\ \hline
Honduras & 17 & 14 & 14 & 16 \\ \hline
Guatemala & 18 & 16 & 12 & 17 \\ \hline
Bolivia & 16 & 15 & 19 & 18 \\ \hline
Nicaragua & 19 & 18 & 16 & 19 \\ \hline
\end{tabular}%
}
\end{table}

A total of 19 countries were studied to get the final ranking. According to Table IV, several Latin American countries are emerging as potential powers for the implementation of artificial intelligence and the reception of investment. Among these countries are Brazil, Mexico, and Chile, which already have a positive reputation in this regard. However, the study also highlighted the emergence of other countries such as Argentina, Colombia, Uruguay, Costa Rica, and Ecuador. 

The study evaluated several factors that contribute to a country's potential for implementing artificial intelligence, such as infrastructure, academic status, and economic status. The results showed that Argentina, Ecuador, and Uruguay have the best infrastructure among emerging powers. Colombia, Argentina, and Costa Rica have the best academic status, while Colombia, Argentina, and the Dominican Republic have the best economic status among emerging powers. 

\begin{figure}[h]
    \centering
    \includegraphics[width=1\linewidth]{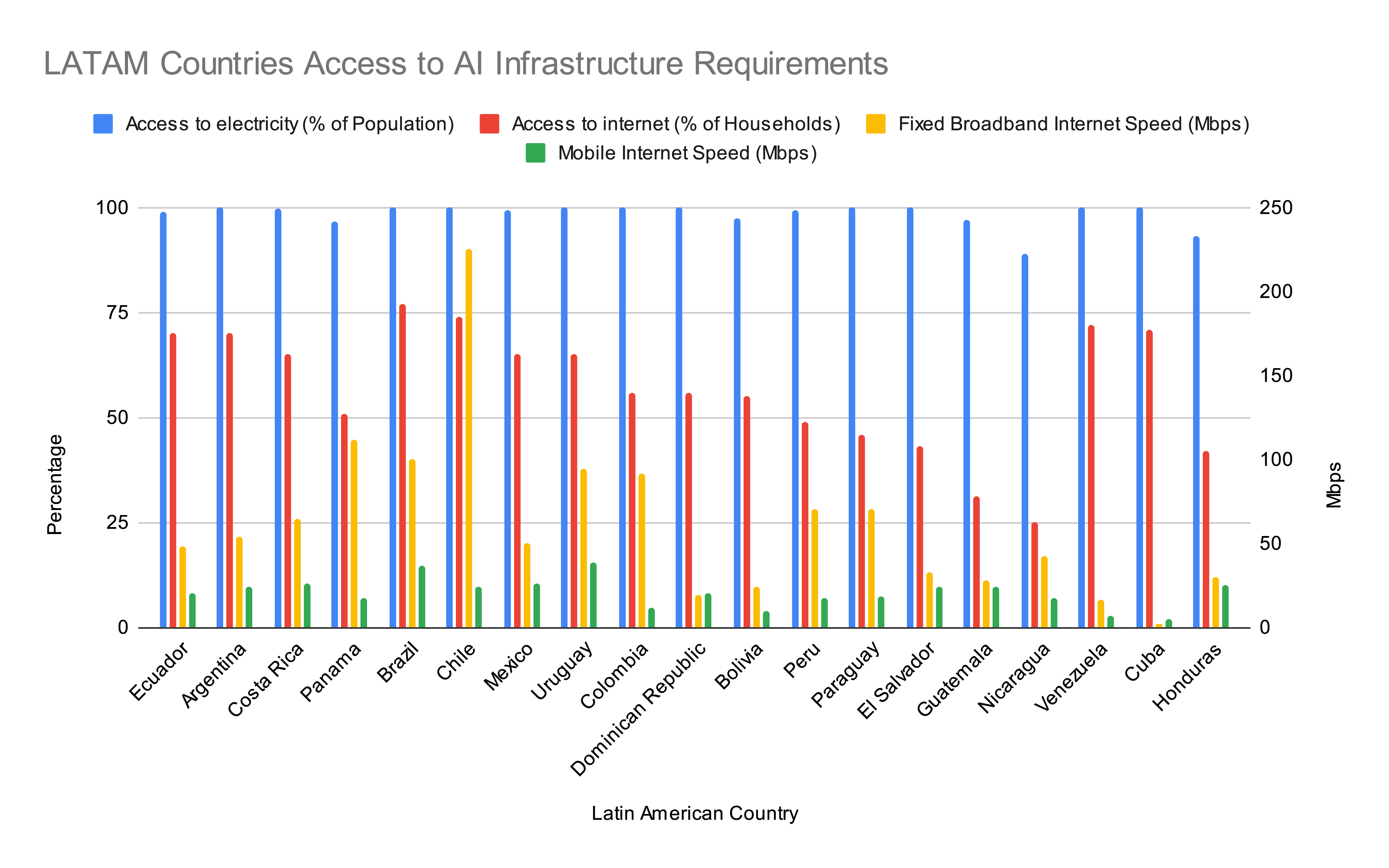}
    \caption{Comparing the percentage of the population who have access to electricity, percentage of households with fixed internet connection, and speed of broadband and mobile internet in Mbps in Latin American countries. Data for electricity access was sourced from the World Bank Group, data for internet access was sourced from the World Bank Group and UNDP, and internet speed data was sourced from Speedtest.}
    \label{fig:Infra-chart}
\end{figure}

Figure 1 demonstrates the current status of basic infrastructure in the Latin American region. Brazil, Chile, and Mexico have the best infrastructure in terms of internet and electricity\cite{WorldBankOpenData2023AccessPopulation}. However, other potential countries such as Panama, Uruguay, and Colombia also have the basic infrastructure necessary to implement artificial intelligence in their territory\cite{WorldBankGroup2022Internet2021}\cite{Speedtest2023InternetWorld}.

High-Performance Computing (HPC) systems are crucial for the proper implementation of AI as they can efficiently process large amounts of data during the training of AI algorithms, aiding research, and increasing productivity and development.

\begin{figure}[h]
    \centering
    \includegraphics[width=1\linewidth]{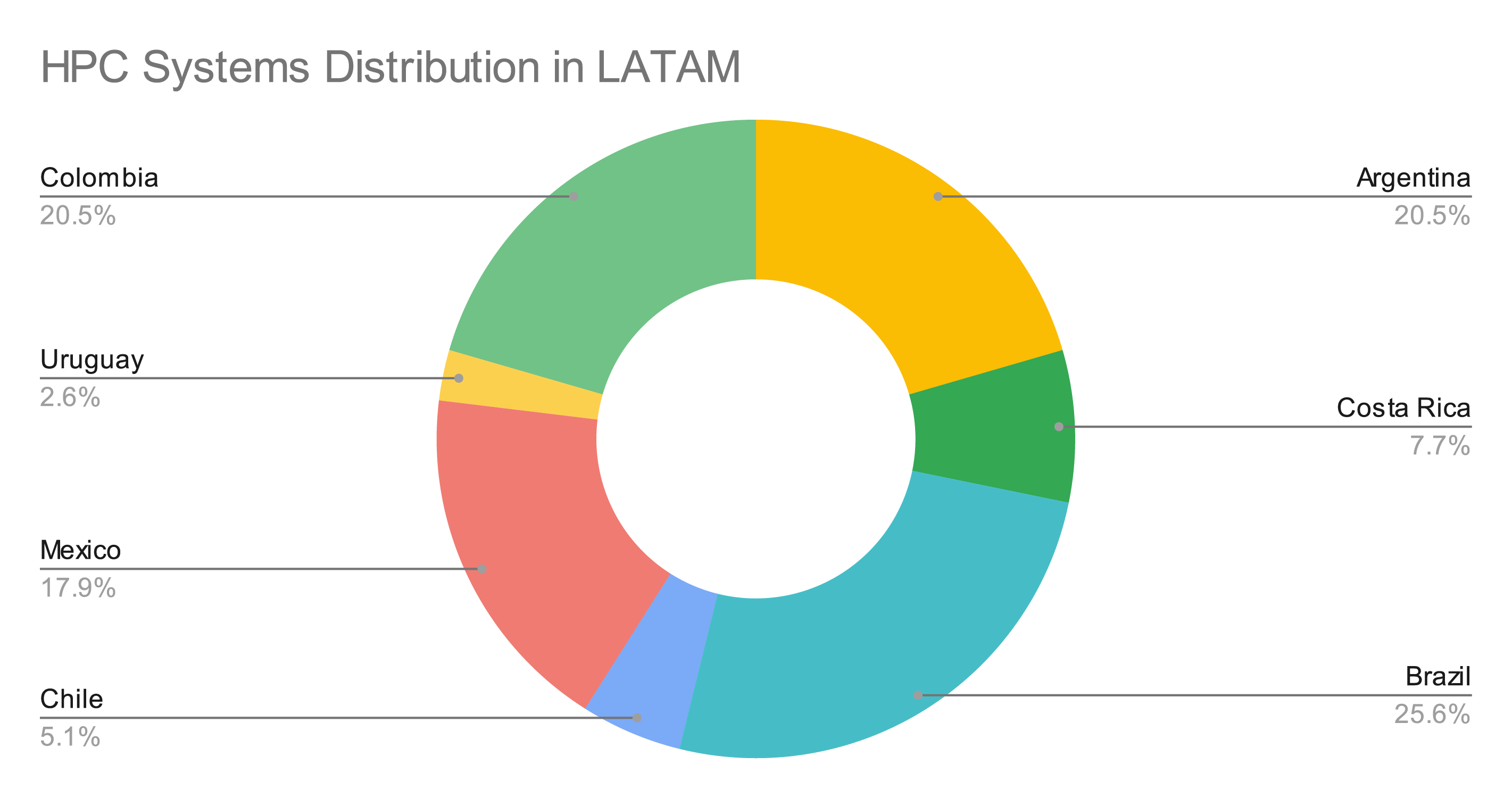}
    \caption{Distribution of the HPC system in Latin American countries in percentage terms. Data was sourced from RISC2 Deliverable 2.1 White Paper on HPC RDI in LATAM}
    \label{fig:HPC-dist}
\end{figure}

Figure 2 shows that Brazil, Chile, and Mexico, again, are leading in AI and technology in Latin America. Uruguay, Costa Rica, and Argentina have the potential to become future AI and technology powers due to their good distribution of HPC systems and efforts to promote investment and implementation of AI\cite{Hafner2021RISC2LATAM}.

\begin{figure}[h]
    \centering
    \includegraphics[width=1\linewidth]{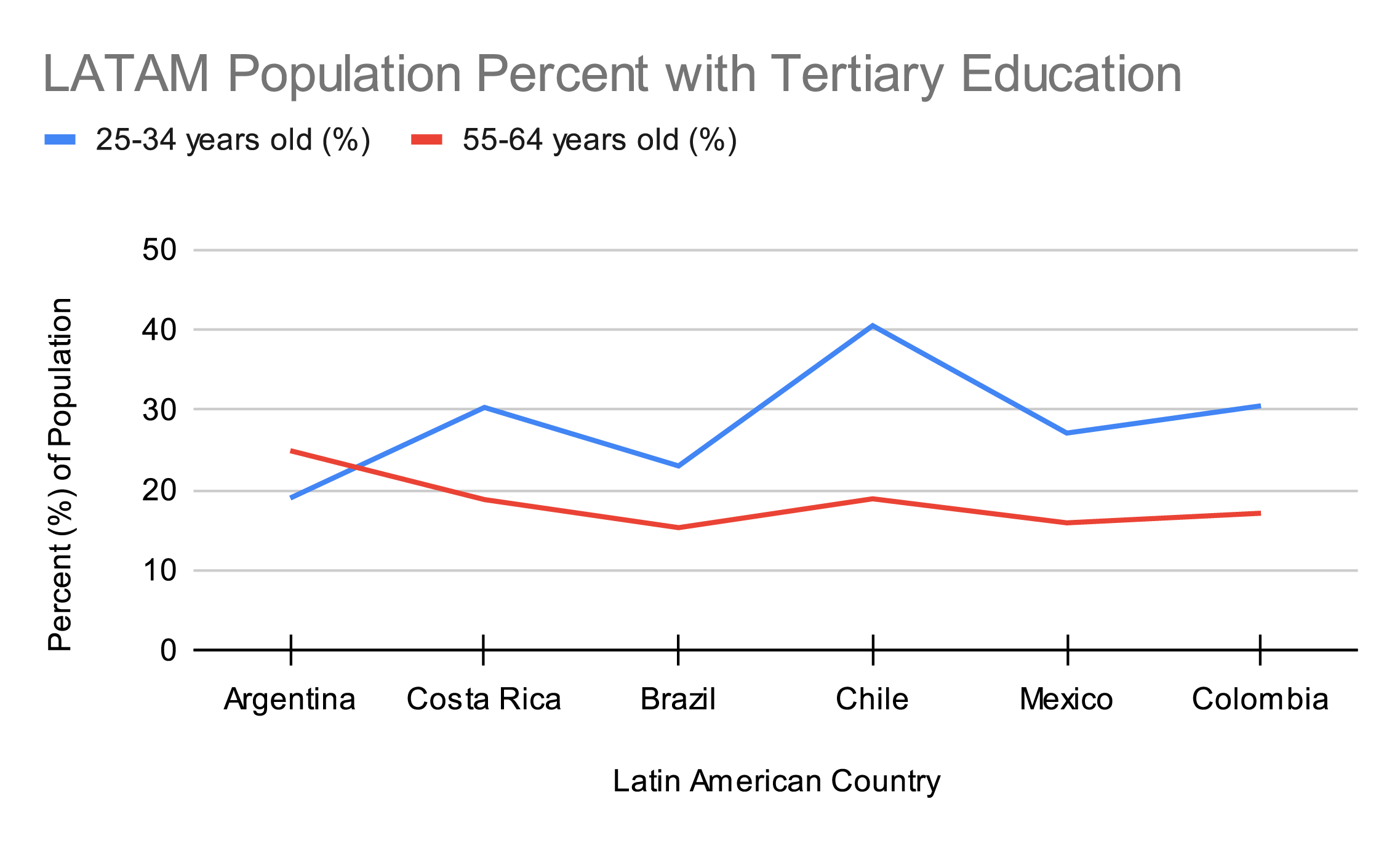}
    \caption{Distribution of the percentage of population with tertiary education by age slots in Latin American countries. Countries with no data available were excluded from the figure. Data sourced from the OECD.}
    \label{fig:Tert-Ed}
\end{figure}

Figures 3 and 4 demonstrate how in the educational field, it can be noted that Brazil, Chile, and Mexico, as in the previous cases, are predominantly, as well as the new possible technological powers of the regions such as Costa Rica, Argentina, and Uruguay, who are beginning to give themselves a place among the best educationally positioned countries in Latin America\cite{QSQuacquarelliSymonds2023QS2023}.

\begin{figure}[h]
    \centering
    \includegraphics[width=1\linewidth]{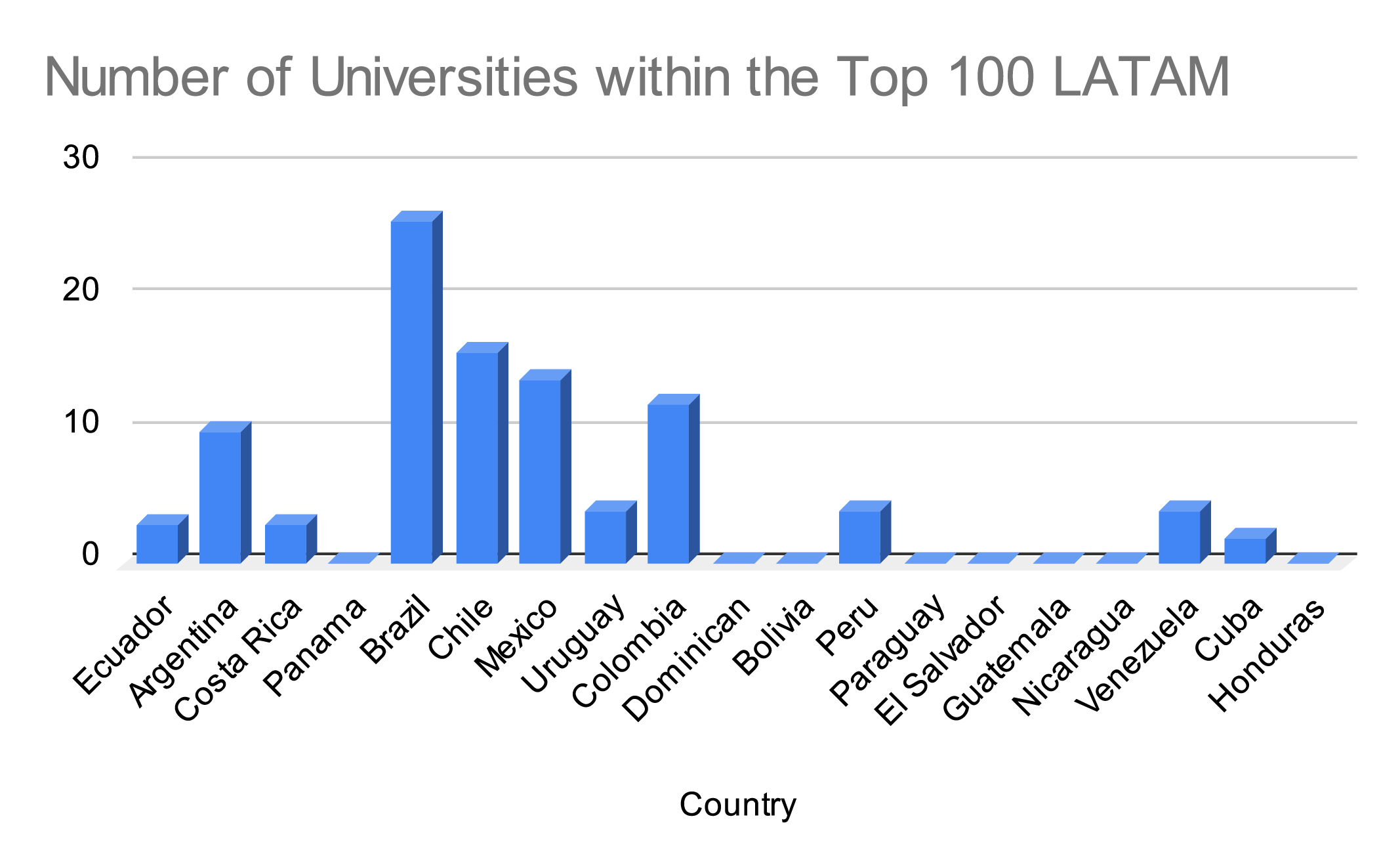}
    \caption{Distribution of the top 100 Latin American Universities. Data sourced from QS TOP UNIVERSITIES.}
    \label{fig:Universities-100}
\end{figure}

Argentina stands out as the only country in Latin America where the older population has a higher level of education than the younger population\cite{OECDPopulationEducation}. This fact is not necessarily negative, as the older population also has a higher level of tertiary education. This means they can participate more actively in society and contribute positively to the country's economy. Additionally, this characteristic of the Argentine population may lead to more informed and responsible decisions by the older generation due to their higher level of education, thus making the country more financially stable. For investment and implementation of AI, the benefits of the older population having more education than the young population include the increase in the digital literacy and competence of the older population which can enable them to use AI. Also, the digital financial and security skills of the older population help them to make better investment decisions and plan for their retirement. Finally, the increase in the social and civic engagement of the older population can foster inter-generational collaboration in the exchange of ideas and perspectives and promote the ethical and inclusive use of AI technologies.

On the other hand, Costa Rica also occupies an important position in terms of population with higher education in Latin America, with its youngest population being the most educated, but at the same time, its older population has the second highest percentage of tertiary education below from Argentina.

\begin{figure}[h]
    \centering
    \includegraphics[width=1\linewidth]{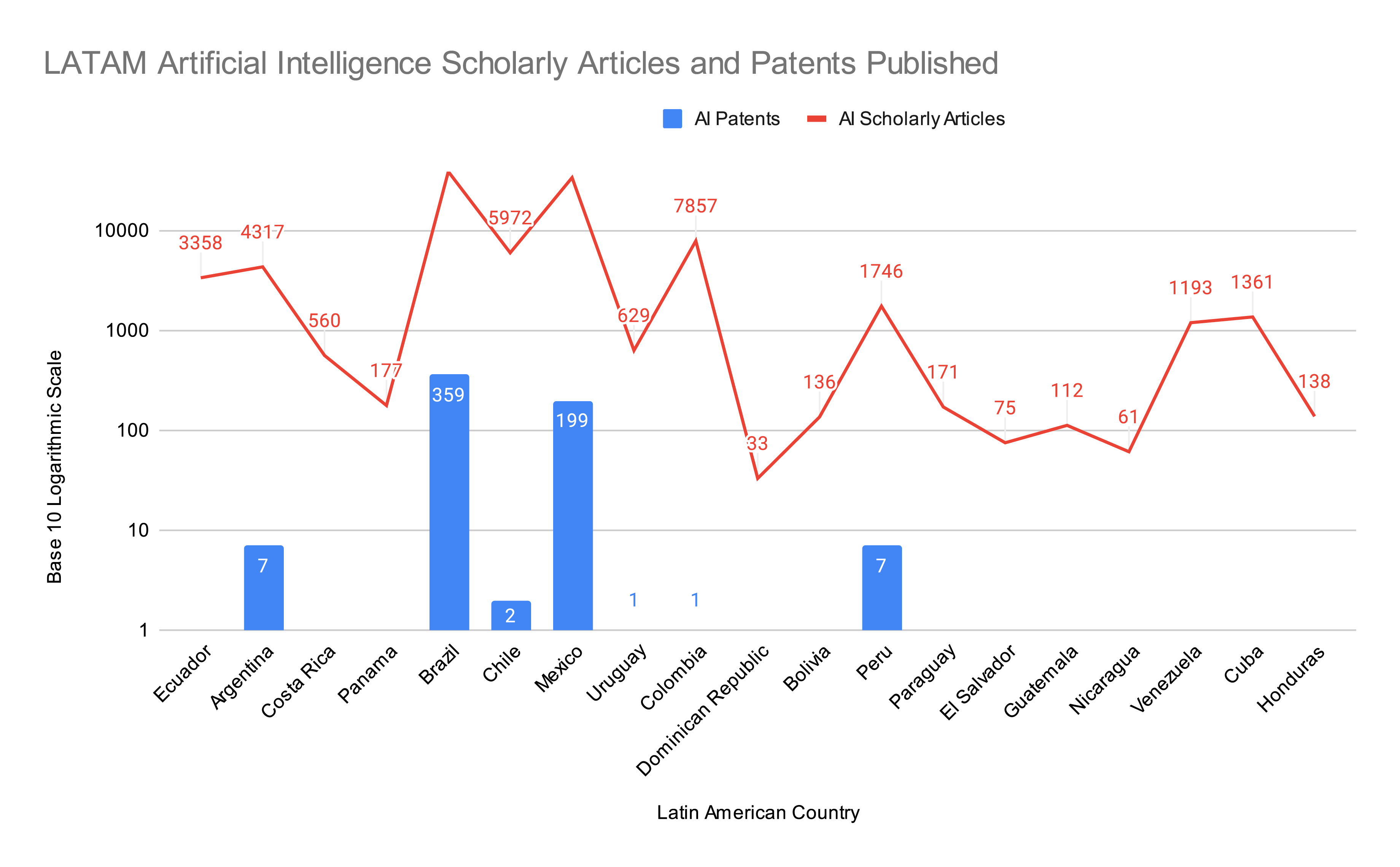}
    \caption{Comparison of patents and scholarly articles published under the AI field in Latin American countries. Data sourced from the Emerging Technology Observatory.}
    \label{fig:latam-publications}
\end{figure}

According to Figure 5, Brazil and Mexico are leading in AI research and development in Latin America. Brazil has published the most articles on AI with 38,775 publications, and Mexico is second with 33757 publications. Additionally, Brazil has 359 granted patents in the AI field, and Mexico has 199, while the rest of Latin America has less than 10 patents\cite{CenterofSecurityandEmergingTechnology2023CountryIntelligence}. These numbers demonstrate the strong commitment of these nations toward AI development and suggest the potential for economic growth and improved quality of life.

Nevertheless, other countries have also begun to increase research on the benefits and applications of artificial intelligence as Colombia, Ecuador, Chile, and Argentina, this will allow different strategies adapted to specific cases to be implemented for both companies' and governments' plans.

It is worth noting that the adoption of artificial intelligence in organizations in Latin America was 52\% in 2021 and 44\% in 2022\cite{Human-CenteredArtificialIntelligence2023Artificial2023}. This trend is reflected in  Fig 6, which provides details on the total estimated value of incoming investments in Latin America and Tech transactions. 

\begin{figure}[h]
    \centering
    \includegraphics[width=1\linewidth]{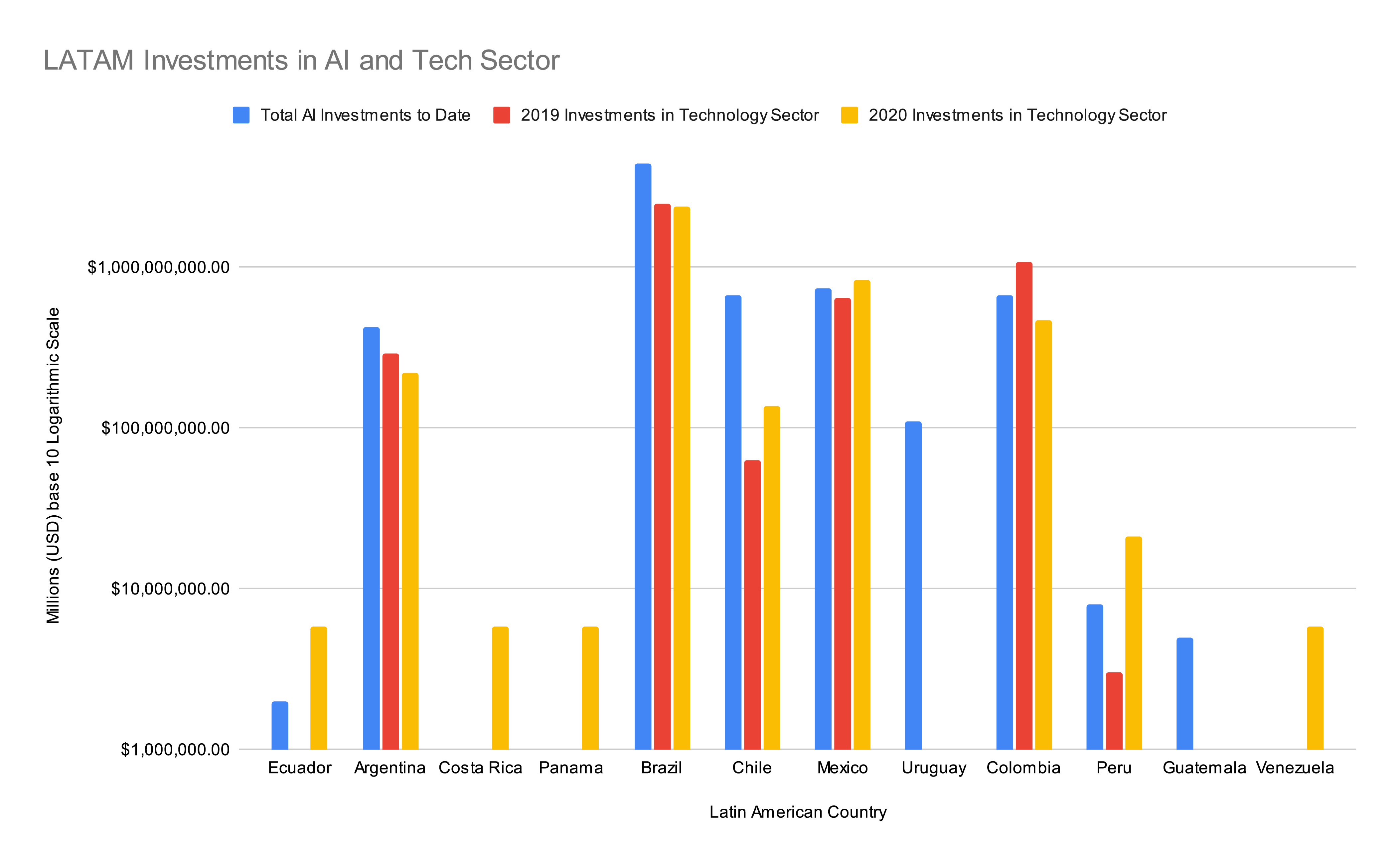}
    \caption{Blue bar demonstrates the estimated value of incoming investments in AI in Latin American countries to date in USD millions. Red and Yellow bars show the distribution of Latin American investments within the AI Field during the years 2019 and 2020 expressed in USD millions. Data sourced from the Emerging Technology Observatory and LAVCA.}
    \label{fig:current-latam-invest}
\end{figure}

It's essential to mention that while some countries have yet to receive investments in AI as depicted in Figures 6, they still have a tremendous potential for growth and advancement in this field. For instance, Costa Rica holds the 3rd most attractive spot in Latin America for mergers and acquisitions (M\&A) investments, with 54\% interest, followed by Peru in 6th place with 47\%, Uruguay in 7th place with 47\%, Argentina in 8th place with 44\%, and Panama in 9th place with 33\%\cite{GarciadePresno2023KPMGSurvey}.

Additionally, the technology sector is one of the most important sectors for investors in Latin America. Between 2019 and 2020, over 8 billion dollars were invested in technology companies, with Brazil being the country with the highest investment. In 2019, Brazil received 2.49 billion dollars, and 2.22 billion in 2020\cite{LAVCA2021LAVCAsAmerica}.

In 2020 in Latin America, the most investment in the technology sector went to the financial industry that applies technological solutions; e-commerce, which is about buying and selling products virtually; and super apps, which are based on applications where users can perform multiple tasks without the need to use another application\cite{LAVCA2021LAVCAsAmerica}.

These large investments have contributed to the economic growth of Latin American countries, achieving the projections of change in the Real GDP of most countries for 2023, 2024, and looking towards 2028 is between 2\%, up to more than 4\% annually. Chile is the only country in the region with a negative projection only in 2023 with -1\%\cite{InternationalMonetaryFund2023WorldRecovery}.
 
\begin{figure}[h]
    \centering
    \includegraphics[width=1\linewidth]{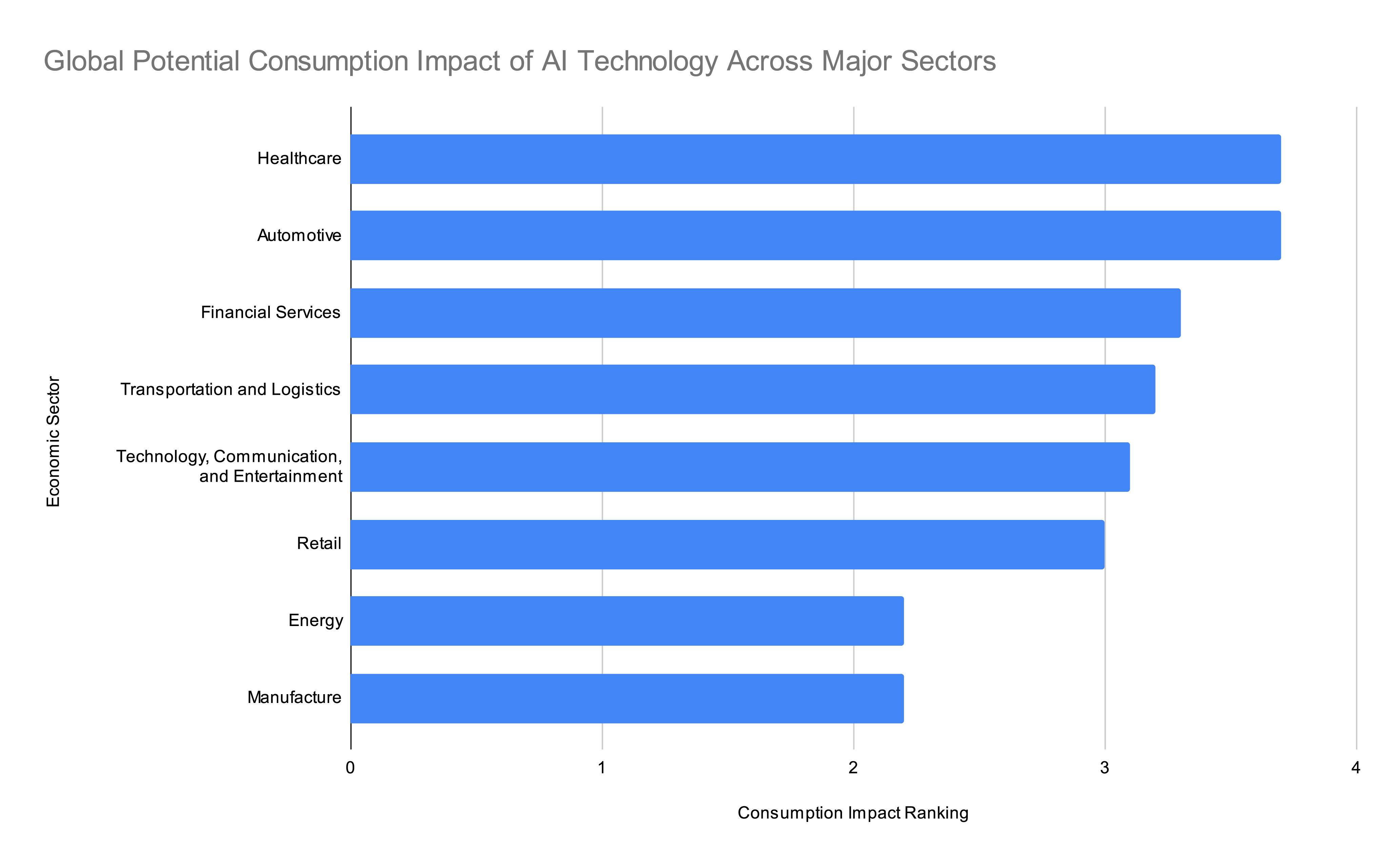}
    \caption{Global Potential Consumption Impact of AI Technology in the Largest Economic Sectors. Impact ranking developed by PWC, which compares the average rate of AI adoption across business sectors in near-term (0-3 years), mid-term (3-7 years), and long-term (7+ years) timelines.}
    \label{fig:consumption-impact}
\end{figure}

GDP PPP per capita is a crucial factor to consider before investing in any particular country. It helps to determine the purchasing power of the population and the level of economic development. This information is valuable to assess whether the population has enough resources to afford a product or service, which is planned to be produced. Moreover, it aids in analyzing whether these resources could be used to invest in advanced technologies. Figure 7 demonstrates the Global potential consumption impact of AI technology in the largest economic sectors.

\begin{figure}[h]
    \centering
    \includegraphics[width=1\linewidth]{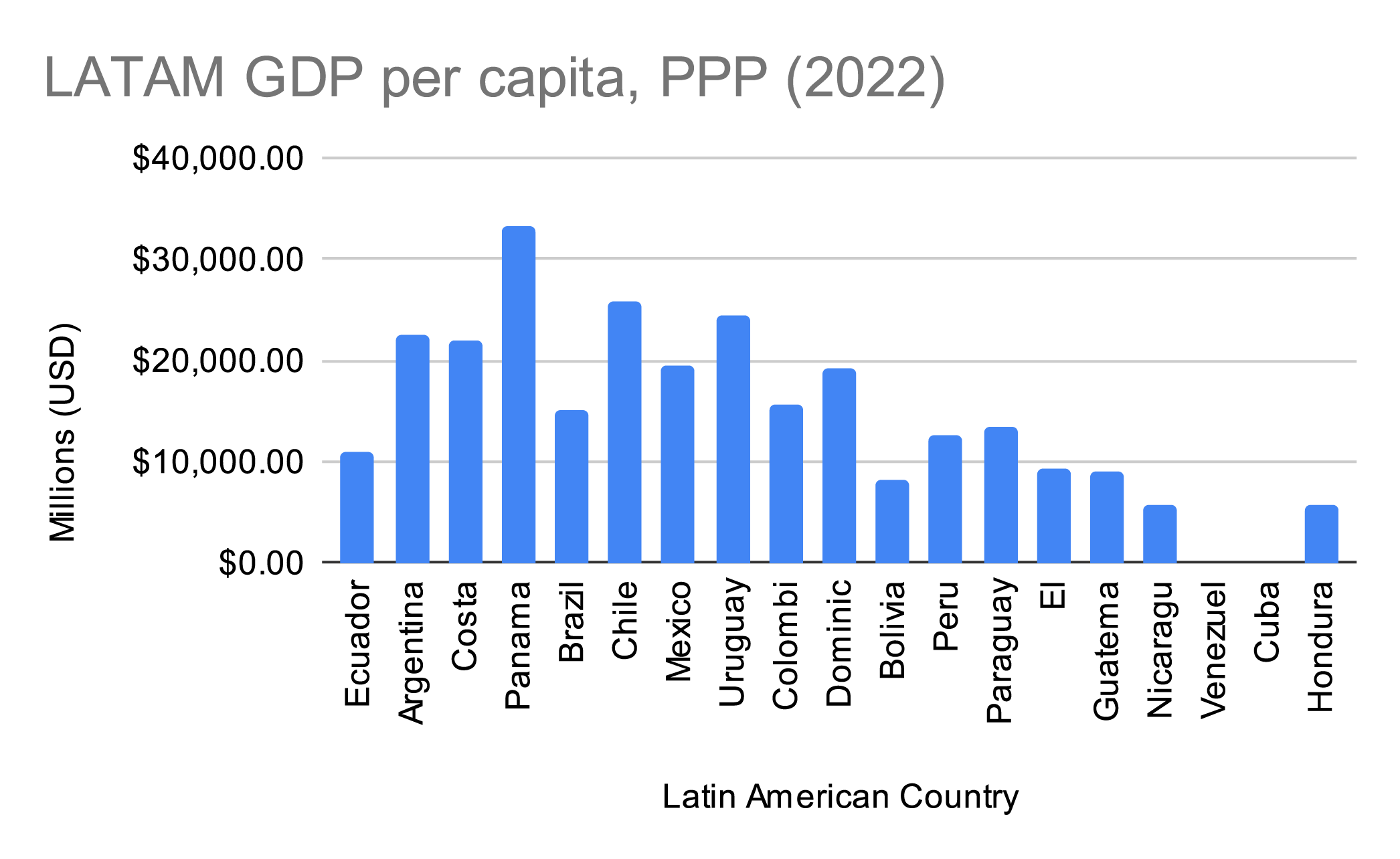}
    \caption{GDP per capita PPP in 2022 in Latin American countries. Data sourced from The Global Economy.}
    \label{fig:gdp}
\end{figure}

According to Figure 8, Panama, Chile, Uruguay, Argentina, and Costa Rica have the highest PPP in Latin America for the year 2022\cite{TheGlobalEconomyGDPRankings}. This, along with the projected growth in Real GDP, could make these countries more attractive for foreign investment in artificial intelligence.

When it comes to investment strategy, salary costs play a vital role. Latin America is a preferred destination for many startups due to its lower salary costs compared to the United States or Europe. For instance, the median salary for roles such as Software Developers, Quality Assurance Analysts, and Testers in the United States in 2022 was \$124,200 per year\cite{USBureauofLaborStatistics2023SoftwareHandbook}. In comparison, as specified in Figure 9, countries like Chile and the Dominican Republic offer higher average salaries that exceed \$60,000 per year for similar roles in computer science. Countries such as Panama, Costa Rica, Ecuador, and Uruguay represent a future potential for the implementation of artificial intelligence, and their average salaries do not exceed \$50,000 per year. This means that startups can reduce their salary costs by more than 50\% compared to the United States\cite{StackOverflow20222022Survey}.

\begin{figure}[h]
    \centering
    \includegraphics[width=1\linewidth]{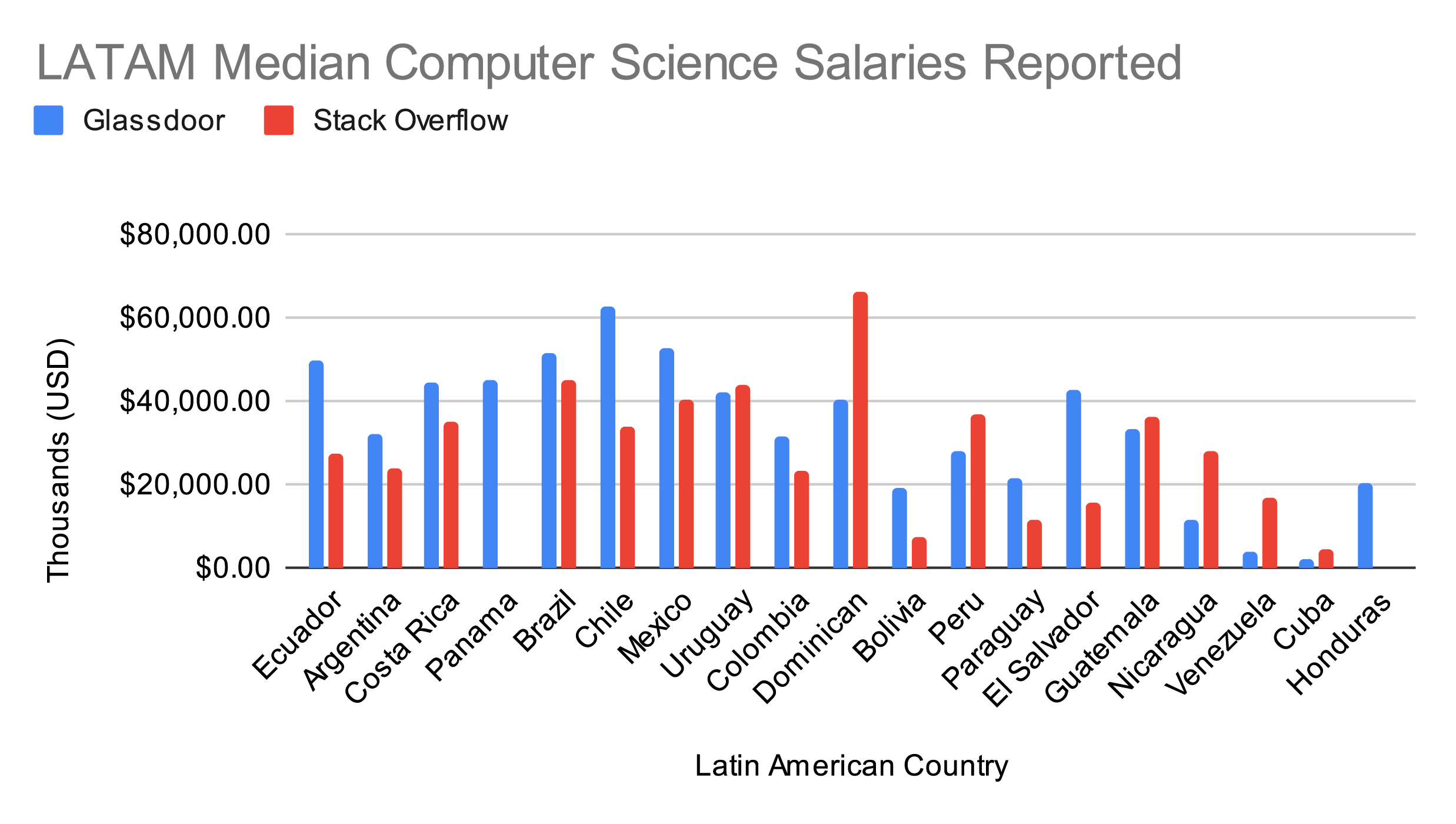}
    \caption{Median Salary for Computer Science roles in Latin American countries. The roles taken into account include Software Engineers, Machine Learning Engineers, Machine Learning Researchers, AI Researchers, Data Engineers, IR Project Managers, and Cyber Security Engineers. Data sourced from Glassdoor and Stack Overflow.}
    \label{fig:cs-salaries}
\end{figure}

It is noteworthy that while the Dominican Republic ranks 7th on the list, it was not included among the emerging powers due to its notable deficiency in the field of education. The study emphasized the need for a balance between the various factors studied to appear as an emerging power in artificial intelligence.

\section{CASE STUDIES}

\subsection{Recovering from Economic Collapse: Triumphs Amidst Uncertainty}

It's interesting to reflect on how some of these countries have managed to overcome economic collapse and achieve full recovery. Despite the risks involved, there are several success stories of companies in Latin American countries that have thrived even in times of uncertainty. It's inspiring to see how these companies have managed to overcome the challenges and achieve growth amid adversity.

\subsubsection{Argentina}

Argentina faced a major economic crisis from 2001 to 2011, due to poor government administration, loss of confidence in the national financial system, and a global economic recession that was affecting many countries at the time\cite{Pagni2012LAKirchner}. Despite this challenging situation, certain companies, such as Mercado Libre, which was established just before the crisis, were able to merge and implement investment strategies to achieve positive results. Mercado Libre started as an online auction marketplace in 1999 and later developed various solutions to benefit Argentina and the rest of Latin America. During Argentina's economic recession, Mercado Libre became the first tech company in the country to be listed on the Nasdaq in 2007\cite{MercadoLibreHistoriaRecorrido}, setting an example for other companies seeking to succeed during difficult times.

\subsubsection{Ecuador}

Ecuador has faced two significant economic crises in the last 25 years. The first one took place in 1998 when the natural disasters caused by El Niño severely affected the country's infrastructure. Additionally, the international oil prices in Ecuador fell by 51\% since 1996, which affected the country's tax revenues, as hydrocarbons accounted for 40\% of the same between 1991 and 2000\cite{Larrea2009CrisisLatina}. However, Ecuador managed to recover from this economic crisis by implementing different financial strategies, such as dollarization, which was able to curb inflation.

Unfortunately, Ecuador, like many other countries, also suffered a financial collapse due to the economic stagnation caused by the coronavirus pandemic in 2020. This resulted in a contraction of 7.8\% in GDP compared to 2019\cite{BancoCentraldelEcuadorLA1998}. However, Ecuador was able to recover from this crisis by implementing various financial and health measures. For instance, they implemented massive vaccination plans that managed to stop the increase in COVID-19 cases\cite{WorldBank2021EcuadorDias}. Additionally, Ecuador diversified its exports to compensate for the decline in exports of goods such as bananas and other products, which caused a reduction in the country's income\cite{LaHora2022LaExportaciones}.

There have been difficult times in Ecuador, but some companies founded during these crises have been able to survive and remain stable. One such example is Vision Fund Bank, established in 1995. In 2019, during its first and only round of financing, it received 6.1 million Euros\cite{CrunchbaseBancoEcuador}. 

Despite facing economic crises in 1998 and the COVID-19 pandemic, Banco VisionFund was able to weather these challenges. It implemented strategic actions such as digitalization of services such as online banking, electronic wallets, and payment networks. It also implemented health strategies such as providing protection and biosafety protocols for its workers. However, in 2020 and 2021, Banco VisionFund’s net income decreased to 591.1 million dollars\cite{BancoVisionFund2020Balance2020} and 552.73 million dollars respectively\cite{BancoVisionFund2021Balance2021}.

Nevertheless, in 2022, Banco VisionFund was able to increase its net income by 310\% compared to the previous year. In 2021, its net profitability was 1,095.36 million dollars, and by 2022, it had increased to 3,400.17 million dollars\cite{BancoVisionFund2022Balance2022}.

\subsubsection{Costa Rica}

During the 80s, Costa Rica faced a significant economic decline due to both internal and external factors. The mismanagement of the State's economic resources, the aftermath of the Second World War, and the rise in hydrocarbon prices were among the primary causes. These factors resulted in a substantial increase in the country's public debt, which rose from 833 million dollars in 1977 to 3.3 billion dollars in 1983. Inflation rates in the country also soared, with an increase from 9.19\% in 1979 to 18.12\% in 1980, 37.06\% in 1981, and 90.12\% in 1982\cite{Garita2006Crisis80}. 

Although the Costa Rican economy was affected by the global recession, increase in international interest rates, deterioration of trade with Central American economies, poor management of public finances, and excessive consumption of Costa Rican imports, there are,  as in the previous examples, organizations that were created during times of uncertainty which were able to demonstrate that good management of funds and a well-implemented strategy can carry a company forward even in difficult times. In this case, this model is CINDE, a non-profit organization created to increase sustainable foreign investment in Costa Rica during the economic decline in 1982. CINDE was able to promote investment in the country, managing to guide and advise more than 400 high-tech companies\cite{CINDE2023Cinde:Solutions}, and, in turn, positioning itself as the number one investment promotion agency. Additionally, they were able to achieve in 2021 an approved budget of \$6.5 million, 77\% through contributions from PROCOMER, COMEX, and ICT, and the other 23\% through their contribution\cite{CINDE2021Impact2021}.

\subsubsection{Panamá}

Panama experienced a severe financial crisis from 1988 to the mid-90s. This crisis resulted from inadequate regulation and supervision of the banking system, corruption, speculation in the financial administration of some banks, and a conflict between the president of that time, Noriega, and the administration of the United States government. The US invasion of Panamanian soil in 1989\cite{teleSUR-pha-JGN-DRL2022Que1989} further worsened the situation, causing the destruction of public and private infrastructure, as well as an increase in the unemployment rate due to the closure of companies because of economic sanctions. The crisis also led to the paralysis of Panamanian economic activity, increased inflation, and a political-economic dependence on the United States. As a result, Panama's GDP contracted by 13\% in 1989 and 9\% in 1990\cite{Abrego2016ConsecuenciasPanama,PanamaViejaEscuelaLaPanama}.

In the case of Panama, we can take Do It Center as a model of how Latin American investment, even in times of adversity, can be a good decision not only from an economic point of view but also from a social and environmental one. Do It Center was created in 1990, months after the US invasion. This company was born as a “supermarket-type” hardware store, which was something new in the country and attracted a lot of attention from new customers. Likewise, the company’s continuous innovation produced that Do It Center was able to bring to the market in times of crisis a new purchasing method in the hardware store market, achieving to date more than 29 operational stores nationwide\cite{DoItCenter2023DoHistoria}.

\section{DISCUSSION}

After analyzing the economic, academic, and infrastructure factors, it's evident that countries like Brazil, Mexico, and Chile have made significant technological advancements. However, this study aims to identify which emerging economies in Latin America have the potential to invest in and implement AI within their territories to benefit their populations while providing significant returns. By sharing knowledge and experiences, developing research projects, and collaborating on AI solutions, these countries can create a productive ecosystem throughout the region. 

While countries like Mexico and Brazil are well-established economies in the region, investing in riskier countries can provide significant advantages. For instance, these country's startups typically have lower evaluations, which translates into a lower initial investment. Additionally, the cost of human resources is often lower in these areas, which can further reduce overhead and increase profitability. Furthermore, investing in riskier countries can result in a higher potential for return on successful investments due to a relatively untapped market with significant growth potential. These factors can make investing in emerging economies in Latin America a smart, strategic move for investors looking to diversify their portfolios, expand their reach, and maximize their profits.

Working collaboratively and investing in Artificial Intelligence (AI) can be advantageous for foreign investors seeking a destination with skilled technology human resources, lower labor costs, access to natural resources, infrastructure, tax, and financial incentives. Latin America offers a lot of potential for foreign investors who view it as a promising investment destination in the future. Countries such as Argentina, Colombia, Uruguay, Ecuador, and Costa Rica are among the Latin American nations with enormous potential for investment and implementation in the future.

Argentina has a highly educated population, with a well-developed academic sector and a high percentage of individuals with higher education. The country has achieved 100\% coverage of electricity and a notable availability of internet. In terms of AI, Argentina is actively conducting research, with over 4,300 articles published and seven patents granted. The country has approximately 25 startups and private companies in the sector, which have received investments amounting to an estimated value of 426 million dollars.

Despite having a low average salary compared to other Latin American countries, Argentina has one of the best GDP PPP, with greater purchasing power than countries such as Mexico, Peru, Costa Rica, and Brazil. The country's emerging economy has great potential for strategic investments that will allow for economic and social profitability.

Furthermore, Argentina currently has a plan to dollarize the economy, which is said to be the only way to end inflation and generate a revaluation of production and salaries in dollars. This plan is similar to the one implemented by Ecuador, which proved to be beneficial for the country.

Colombia emerges as a promising hub for artificial intelligence (AI) investment and implementation, boasting robust infrastructure, 100\% electricity access, and favorable internet speeds. Hosting 20.5\% of High-Performance Computing (HPC) systems in Latin America, the nation stands out. In academia, Colombia holds the fourth-largest population aged 25-34 with higher education, featuring 12 of the top 100 regional universities and ranking third in AI article publications. Economically, it demonstrates allure with an estimated \$676 million in AI investments, 16 AI startups, and cost-effective computer science roles, offering an average annual salary of \$31,765. The government's pro-AI strategies align with a 2021 GDP per capita PPP of \$14,648.59 and projected real GDP growth rates of 1\% (2023), 1.9\% (2024), and 3.3\% (2028).

Uruguay secures a position in the top 6 countries in our Latin American Artificial Intelligence ranking for compelling reasons. With a robust infrastructure, the nation boasts 100\% electricity access, 65\% internet coverage, and penetration rates of 67\% in urban and 38\% in rural areas. Uruguay showcases high-speed internet connections, averaging 94.41 Mbps in fixed broadband and 38.59 Mbps in mobile, along with 2.6\% of HPC systems in Latin America. Academically, it houses four of the region's top 100 universities, publishing 629 AI articles and securing one patent. Financially attractive, Uruguay offers a competitive average annual salary of \$42,000 for Computer Science roles. Supported by government strategies, the country's GDP per capita PPP of \$22,800 signifies strong purchasing power, with projected Real GDP growth rates of 2\% (2023), 2.9\% (2024), and 2.2\% (2028).

Ecuador stands out as a key player in Latin America, ranking 4th in AI infrastructure. With a robust 70\% nationwide internet access, including 77\% in urban and 60\% in rural areas, Ecuador fosters an ideal environment for AI implementation. Boasting 98.8\% electricity access and stable internet speeds, the country is well-positioned for AI development. Academically, Ecuador excels with 3,358 AI articles and three universities in the top 100 in Latin America. With an estimated \$2 million AI investment, Ecuador aligns with the regional average, offering a competitive \$38,000 annual salary for computer science roles. The IMF projects steady real GDP growth of 2.9\% (2023), 2.8\% (2024), and 2.8\% thereafter. Ecuador's economic transformation post-dollarization, marked by reduced inflation and positive GDP growth, positions it as a notable investment destination, ranking third in foreign direct investment increase in 2019 in the region.

Costa Rica is one of the Latin American countries that made it to the list. Despite being a smaller economy, the country boasts of a good infrastructure, with 99.9\% access to electricity and 65\% internet access. It also has 7.7\% of the HPC systems in Latin America. In terms of academic level, Costa Rica has a population of 30.3\% between 25 and 34 years old with higher education. Additionally, 3 out of the 100 best universities in Latin America are located in Costa Rica, where over 560 articles within the field of artificial intelligence have been published. 

On the economic side, salaries within Computer Science roles in Costa Rica are at an average point between salaries within these roles in Latin America, which is approximately \$39,000. The country had a PPP GDP per capita of \$21,200 and expects a projected increase in its Real GDP for 2023 of 2.7\%, 3.2\% in 2024, and 3.2\% in 2028.

\section{CONCLUSIONS}

Latin America is a region with immense potential for development, which can be boosted by foreign investment and strategic implementation of artificial intelligence. Investing in AI can significantly benefit various sectors of the region's economy by lowering costs, increasing productivity, and promoting technological advancements. 

The imbalance of these factors can increase the investment risk and make the implementation of artificial intelligence more complicated. Therefore, countries need to focus on improving their infrastructure, academic status, and economic status to become a potential destination for investment and the implementation of artificial intelligence. 

In summary, the emergence of several Latin American countries as potential powers for implementing artificial intelligence and the reception of investment is a positive development. However, it is necessary to focus on improving the various factors that contribute to this potential to ensure a balanced and sustainable approach.

The Latin American region has already been receiving foreign investment and implementing strategies for the proper use of artificial intelligence technology. However, compared to other regions of the world, the adoption of these technologies is still low. Therefore, it is crucial to continue promoting investment and adoption of artificial intelligence to strengthen initiatives and collaborations in this field. This will contribute to the growth of the countries' economies.

Latin America has the potential to advance its development through the adoption of artificial intelligence. Therefore, governments, companies, and universities should collaborate to create an ecosystem that is conducive to technological innovation, adoption, and investment.

\bibliographystyle{IEEEtran}
\bibliography{references}

\end{document}